\documentclass{ckm}                 
\def\gsim{{\mathrel{\raise2pt\hbox to 8pt{\raise -5pt\hbox{$\sim$}\hss{$>$}}}}}
\def\rsim{{\mathrel{\raise2pt\hbox to 8pt{\raise -5pt\hbox{$\sim$}\hss{$>$}}}}}
\def\lsim{{\mathrel{\raise2pt\hbox to 8pt{\raise -5pt\hbox{$\sim$}\hss{$<$}}}}}

\confname{Workshop on the CKM Unitarity Triangle, IPPP Durham, April 2003}

\title{Estimates of Light Quark Masses from Lattice QCD and QCD Sum rules}

\author{Rajan Gupta\addressmark{a} 
        \thanks{It is a pleasure to thank Patricia Ball, Jonathan Flynn, and Laurent
Lellouch for the opportunity to present this review, and to Tanmoy
Bhattacharya and Kim Maltman for many discussions. I also thank 
S. Aoki, T. Kaneko, C. Bernard, V. Lubicz, H. Wittig, and G. Schierholz for 
discussions of their calculations.  The work of
R.G. was, in part, supported by DOE grant KA-04-01010-E161.}
       }

\address[a]{Theoretical Division, Los Alamos National Lab, Los Alamos,
         New Mexico 87545, USA}

\begin{document}

\begin{abstract}
This talk reviews the progress made in the determination of the light
quark masses using lattice QCD and QCD sum rules. Based on preliminary
calculations with three flavors of dynamical quarks, the lattice
estimate is $m_s = 75(15)$ MeV, a tantalizingly low value. On the
other hand the leading estimates from scalar and pseudo-scalar sum rules
are $99(16)$ and $100(12)$ MeV respectively. The $\tau$-decay sum
rule estimates depend very sensitively on the value of $|V_{us}|$. The
central values from different analyses lie in the range $115-120$ MeV
if unitarity of CKM matrix is imposed, and in the range $100-105$ MeV
if the Particle Data Group values for $|V_{us}|$ are used. I also give
my reasons for why the lattice result is not yet in conflict with
rigorous lower bounds from sum rule analyses.
\end{abstract}

\maketitle


\section{Introduction}
\label{sec:intro}
Quark masses are important fundamental parameters of the standard
model.  Our ability to extract them from first principle calculations
of QCD will signal the onset of quantitative control over the
non-perturbative aspects of QCD.  In this talk I will summarize the
current status of the extraction of light quark masses.  All results
for quark masses will be in the $\overline {\rm MS}$ scheme at scale 2 GeV. Most
of the time will be devoted to a review of the lattice results. A
brief summary of sum-rule calculations will also be presented. A more
detailed version of this review is being prepared in collaboration
with Tanmoy Bhattacharya and Kim Maltman~\cite{qm:revBGM:03}.

Of the light quark masses, I will concentrate entirely on the strange
quark mass.  The reason is that, at present, estimates of the
ratios from chiral perturbation theory
\begin{eqnarray}
\frac{2 m_s}{m_u + m_d} &=& 24.4(1.5) \nonumber \\
\frac{m_u}{m_d} &=& 0.553(43)
\label{eq:cptratios}
\end{eqnarray}
are more accurate than the lattice results. Using them to extract
$m_u$ and $m_d$, once $m_s$ is known, avoids some of the uncertainties
due to chiral extrapolations in lattice calculations, and of ignoring
electromagnetic effects in the simulations.

There is a recent result on the charm quark mass that deserves
mention. I mention this to also highlight the fact that current
lattice calculations indicate that the the whole interval $m_{u,d} -
m_c$ can be handled by the techniques used for light
quarks. Consequently, charm quark can be simulated on the lattice with
small discretization errors. Following this approach, the ALPHA
collaboration has presented, in the quenched approximation, the
ratio~\cite{alpha:mc:02}
\begin{equation}
\frac{m_c}{m_s} = 12.0(5) \,.
\label{eq:mcmsratio}
\end{equation}
The advantage of their calculation, based on the Schr\"odinger
functional approach and using the fully $O(a)$ improved theory, is
that they convert lattice masses to renormalization group invariant
masses using a non-perturbative method, thus avoiding the problem of
the perturbative determination of the scale dependent renormalization
constant $Z_m^0 $ at scales $1/a =2-4 $ GeV and the subsequent
matching of the lattice to a continuum regularization scheme at these
scales. They find $m_c^{\overline {\rm MS}}(2 {\rm GeV})= 1.301(34)$ GeV,
which compares well with the recent estimate $m_c^{\overline {\rm MS}}(2
{\rm GeV})= 1.26(4)(12)$ GeV by the SPQcdR
collaboration~\cite{ROME:mc:01}.

\section{Lattice QCD}
\label{sec:LQCD}

A self-consistent determination of quark masses using lattice QCD will
be equivalent to the validation of QCD as the correct theory of strong
interactions. In ideal lattice QCD simulations we need to dial six
input parameters in the generation of background gauge configurations
and then calculate quark propagators on them. These are the gauge
coupling $g$, and the masses for up, down, strange, charm, and bottom
quarks. The top quark is neglected because it is too heavy and too
short lived. If we had petaflop scale computers we could carry out
simuations with three light and two heavy flavors of dynamical quarks
with masses tuned to roughly their physical values $m_u$, $m_d$,
$m_s$, $m_c$, and $m_b$. The inclusion of heavier quarks, charm and
bottom, in the generation of background gauge configurations, however,
are unlikely to significantly affect the vacuum structure and can
therefore be safely neglected in the update for $Q^2$ much less than
$9$ (or $100$) GeV${}^2$. To study their interactions, charm and bottom
quarks are incorporated at the stage of calculating quark propagators
on these background gauge configurations. Thus simulations with 
three light dynamical flavors are the goal of lattice calculations.

Correlation functions in Euclidean space-time, from which various
properties of hadrons are extracted, are constructed by tying together
these quark propagators and gauge link variables in appropriate
combinations. For example the hadronic spectrum and associated decay
constants are extracted from two-point correlation functions with the
appropriate quantum numbers. The masses of hadrons, at least of the
stable ones and of those with very narrow widths, are determined from
the rate of fall-off of these correlation functions at large Euclidean
time. If QCD is the correct theory, these should agree with
experiments up to electromagnetic corrections which are neglected in 
current simulations.

Unfortunately, today's computers are not powerful enough to carry out
calculations with physical values of up and down quark
masses. Instead, quarks heavier than their physical values, $i.e.$ in
the range $m_s/8 - m_s$, have been studied. Also, isospin breaking and
electromagnetic effects have been neglected, $i.e.$, simulations have
been done with $m_u=m_d$ and electric charge turned off.  From these
simulations one extracts physical results by extrapolation in the up
and down quark masses. The calculated hadron masses or decay constants
are expressed as an expansion in quark masses using expressions
derived from chiral perturbation theory. Once the coefficients of
these chiral expansions, which are related to the low energy constants
in the chiral Lagrangian, are determined then one has an overcomplete
set of relations (because the number of hadronic observables are much
larger than the input parameters) between experimentally measurable
quantities and quark masses. Using these relations, extrapolations to
the physical values of hadron masses or decay constants specify the
physical quark masses. A self-consistent determination of the quark
masses in terms of hadron masses or vice versa would validate QCD.

The success of this program requires that three conditions be
met. First, the simulations should be done with three flavors of
dynamical quarks and the input masses for all three quarks (both in
the update and in the construction of external quark propagators,
$i.e.$, sea and valence quarks) should be light enough to lie within
the range of validity of the $O(p^4)$ chiral
Lagrangian~\cite{sharpe:pqqcd:00,sharpe:pqqcd:01,Sharpe:pqqcd:03}. This
condition guarantees that the extracted chiral coefficients are the
same as in QCD and reliable. Second, the simulations should be done at
lattice scales small enough that discretization errors can be
neglected or can be removed by a reliable extrapolation of the data to
$a = 0$. Third, simulations should be done on large enough lattices so
that finite volume effects are negligible.

There are two ways in which the renormalized quark mass at scale
$\mu$ is defined using lattice simulations done at scale $1/a$:
\begin{eqnarray*}
m_R(\mu) &=& Z_m(\mu,a) \ m(a)  \\
(m_1 + m_2)_R(\mu) &=& \frac{Z_A}{Z_P(\mu, a)} \frac 
		{\langle 0 \mid \partial_4 A_4 (t) J(0) \mid 0 \rangle}
		{\langle 0 \mid P_a (t) J(0) \mid 0 \rangle} \,.
\label{eq:mdefn}
\end{eqnarray*}
The first method is based on the vector Ward identity and $m(a)$ is
the bare lattice mass. The second method exploits the axial Ward
identity and uses two-point correlation functions with source $J$
having pseudoscalar quantum numbers.  For lattice
formulations with an exact chiral symmetry, $e.g.$ staggered fermions,
the two methods are identical. The connection between results obtained
in the lattice regularization scheme and some continuum scheme like
$\overline {\rm MS}$ used by phenomenologists is contained in the
$Z's$. Their calculation introduces an additional source of systematic
error in all quantities whose renormalization constants are different
in the two schemes; spectral quantities (hadron masses) which do not
get renormalized are an exception. I later discuss the quantitative
effect on quark masses of the renormalization factors needed to
connect lattice results to those in the $\overline {\rm MS}$ scheme.

In this talk I will analyze the state-of-the-art lattice data and 
discuss their reliability with respect to the following sources of 
systematic errors.
\begin{itemize}
\item
The number and masses of dynamical quarks used in the update of gauge 
configurations.
\item
Chiral extrapolations to physical quark masses.
\item
Continuum extrapolations to $a=0$.
\item
The uncertainty in the calculation of the renormalization constant. In
particular I will discuss the difference in estimates of masses between using 1-loop
perturbative estimate, the one obtained in the RI/MOM scheme, and the
fully non-perturbative one using the Schr\"odinger functional method.
\end{itemize}


\section{State-of-the-art quenched results}
\label{sec:discussQ}

The state-of-the-art quenched results are summarized in
Table~\ref{tab:mq}. Two recent results deserve some elaboration (I
have not included results from domain wall~\cite{Dawson:mq0:03} or
overlap~\cite{Hernandez:mq0:02} fermions as they are still preliminary
and do not include a continuum extrapolation).

{\it (i) Results from the SPQcdR collaboration}~\cite{SPQcdR:mq:02}
supercede all previous estimates from the ROME
group~\cite{ROME:mq:98,ROME:mq:00}, which is why the latter are not
included in Table~\ref{tab:mq}. The new calculations improve on
previous results in the following ways. (i) The SPQcdR calculations
have been done using the non-perturbative $O(a)$ improved
Sheikholeslami-Wohlert action at four values of the coupling,
$\beta=6.0$, $6.2$, $6.4$ and $6.45$. Over this range the lattice
scale changes roughly by a factor of two ($0.1 \to 0.051$ fermi), so a
reliable extrapolation to the continuum limit has been carried
out. (ii) The N${}^3$LO (4-loop) relation is used to connect the
RI/MOM scheme to the $\overline {\rm MS}$ scheme for the
renormalization constants, as well as in the running of the masses to
the final scale $2$ GeV. (iii) Results using both the vector and axial
Ward identity method have been computed and compared.

\begin{table*}
\begin{center}
\noindent
\setlength{\tabcolsep}{4.2pt}
\begin{tabular}{|l|l|c|c|c|c|}
\hline
                          &  Action    & $\bar m$        & $m_s(M_K)$  & $m_s(M_\phi)$& scale $1/a$    \\
                          &  Renorm.   & $=(m_u+m_d)/2$  &             &              &                \\
\hline
JLQCD                     &  Staggered & $4.23(29)$      & $106(7)$    & $129(12)$    & $M_\rho$       \\
(1999)\cite{JLQCD:mq:99}  &  RI/MOM    &                 &             &              &               \\
CPPACS                    &  Wilson    & $4.57(18)$      & $116(3)$    & $144(6)$     & $M_\rho$       \\
(1999)\cite{CPPACS:mq:99} &  1-loop TI &                 &             &              &                \\
CP-PACS                   & Iwasaki+SW & $4.37^{+13}_{-16}$& $111^{+3}_{-4}$ & $132^{+4}_{-5}$    & $M_\rho$       \\
(2000)\cite{CPPACS:mq:00E}& 1-loop TI  &                 &             &              &                \\
ALPHA-UKQCD               & O(a) SW    &                 & $ 97(4)$    &              & $f_K$          \\
(1999) \cite{ALPHA:mq:99} & SF         &                 &             &              &                \\
QCDSF                     & O(a) SW    & $4.4(2)$        & $105(4)$    &              & $r_0 $         \\
(1999)\cite{QCDSF:mq:99}  & SF         &                 &             &              &                \\
QCDSF                     & Wilson     & $3.8(6)$        & $87(15)$    &              & $r_0 $         \\
(1999)\cite{QCDSF:mq:99}  & RI/MOM     &                 &             &              &                \\
SPQcdR                    & O(a) SW    & $4.4(1)(4)$     & $106(2)(8)$ &              & $r_0 $         \\
(2002)\cite{SPQcdR:mq:02} & RI/MOM     &                 &             &              &                \\
\hline
\end{tabular}
\caption{State-of-the-art quenched results for quark masses. The
labels for the action used are: $O(a)$ SW for non-perturbative $O(a)$
improved Sheikhholeslami-Wohlert (SW) fermion action, and Iwasaki for
an improved gauge action. The renormalization factors are calculated
using the 1-loop perturbation theory with Tadpole Improvement (TI) or
the non-perturbative RI/MOM and Schrodinger functional (SF)
schemes. The quantity used to set the lattice spacing $a$ is listed in
the last column with $r_0=0.5$ fermi obtained from the static force relation
$r \partial V(r)/\partial r |_{r=r_0} = 1.65$~\cite{Sommer:r0:94}.  All estimates are
based on extrapolation to the continuum limit. $m_s(M_K)$ and
$m_s(M_\phi)$ refer to the strange quark mass extracted using $M_K$ or
$M_\phi$ to fix it.}
\label{tab:mq}
\end{center}
\end{table*}
%

{\it (ii) The QCDSF collaboration} is in the process of updating their 1999
estimates of the strange quark mass~\cite{QCDSF:mq:99}. They use the
same methodology and the same values of coupling, $\beta=6.0$, $6.2$,
and $6.4$, as the SPQcdR collaboration~\cite{SPQcdR:mq:02}, so their
results will provide a detailed consistency check. Their latest
unpublished quenched estimate is $r_0 m_s^{RGI}=0.341(2)$ which
translates to $m_s^{\overline {\rm MS}}\approx 100$ MeV.

{\bf Discussion:} The estimates in Table~\ref{tab:mq} show a wide
range of values that seem to depend on the lattice action, the
quantity used to set the scale and the quark masses, and the
renormalization constant. What I would like to emphasize is that this
spread does not imply that the lattice calculations are in
conflict, simply that in the quenched approximation one does not
expect consistent results and the spread is a manifestation of that.

Within the lattice community it has been known for some time that, in
the quenched approximation, there is a roughly $10\%$ variation in
estimates of the quark masses depending on the quantity used to set
the lattice scale $a$ ($r_0$ or $M_\rho$ or $f_K$ or $M_N$ etc.). This
was quantitatively demonstrated by Wittig at LATTICE
2002~\cite{wittig:rev:lat02} who converted the results from the four
best simulations by SPQcdR, JLQCD, CP-PACS, and ALPHA-UKQCD
collaborations to a common scale set by $r_0$. The result, shown in
Table~\ref{tab:wittig}, is that these four estimates of $m_s$, when
extracted using a common scale setting quantity, i.e., evaluating
$m_s/r_0$ in the continuum limit, show a much smaller variation and
agree within errors. This indicates that the dependence on the fermion
action, fitting procedures, statistics, and renormalization constants
(perturbative versus non-perturbative) used in the calculations are
much smaller effects.

The other lessons we have learned from quenched simulations are:
\begin{itemize}
\item
Quenched simulations do not give consistent estimates of quark
masses. Estimates depend on the hadronic states used to set the quark
masses. For example $m_s$ set using $M_K$ differs by $15-20\%$ from
that set using $M_{K^*}$ or $M_{\phi}$ when the scale is set by
$M_\rho$. Thus, quark masses are sensitive probes of the effects of
dynamical quarks.
\item
With non-perturbative results for $Z's$ in hand we can evaluate how
well 1-loop tadpole improved perturbation theory works to convert
lattice results to $\overline {\rm MS}$ scheme. We find that 1-loop
estimates work to within $5\%$ for VWI method, $i.e.$ $Z_m$ for Wilson
like fermions, and at about $10\%$ for the AWI method . Given that the
rest of the errors, once a common scale setting quantity is used, are
of this order, the collapse of results in Table~\ref{tab:wittig} to a
roughly common value is not surprising.
\end{itemize}
The bottom line is that quenched simulations have allowed us to refine
the numerical methods, and to understand and quantify all other
sources of errors to within $5\%$. So removing this approximation
becomes the next step in obtaining precise estimates.

\begin{table*}
\begin{center}
\setlength{\tabcolsep}{4pt}
\begin{tabular}{|l|l|l|l|l|}
\hline
\multicolumn{1}{|c|}{Ref.}&
\multicolumn{1} {c|}{$m_s(Q')$}&
\multicolumn{1} {c|}{$Q'$}&
\multicolumn{1} {c|}{$F$}&
\multicolumn{1} {c|}{$m_s(r_0)$}  \\
\hline       	     
JLQCD       & $106(7)$           & $M_\rho$   & $0.90(1)$  & $95(6)$         \\
CP-PACS     & $114(2)(^{+6}_{-3})$     & $M_\rho$   & $0.86(2)$  & $98(2)(^{+6}_{-3})$   \\
SPQcdR      & $106(2)(8)$        & $M_{K^*}$  & $0.87(3)$  & $92(2)(7)$      \\
ALPHA-UKQCD & $\phantom{1}97(4)$ & $f_K$      & $1.02(2)$  & $99(4)$         \\
\hline    
\end{tabular}
\end{center}
\caption{Comparison by Wittig of $m_s(\overline {\rm MS}, {\rm 2\ GeV}, M_K)$ from 
quenched simulations by the JLQCD, CP-PACS, SPQcdR, and ALPHA-UKQCD collaborations. Given are 
the original estimates and the scale setting quantity $Q'$, conversion factor 
$F$ from $Q'$ to $r_0$, and the converted mass.}
\label{tab:wittig}
\end{table*}
%

\begin{table*}
\begin{center}
\noindent
\setlength{\tabcolsep}{4.2pt}
\begin{tabular}{|l|l|c|c|c|c|}
\hline
                          &  Action    & $\bar m$               & $m_s(M_K)$     & $m_s(M_\phi)$   & scale $1/a$    \\
                          &  Renorm    &                        &                &                 &  (GeV)         \\
\hline
JLQCD                     & Wilson+SW  & $3.22(4)$              & $84.5(1.1)$    & $96.4(2.2)$     & $M_\rho$  \\
(2002)\cite{JLQCD:mq2:02} & 1-loop TI  &                        &                &                 & (2.22)  \\
CP-PACS                   & Iwasaki+SW & $3.45^{+0.14}_{-0.20}$ & $89^{+3}_{-6}$ & $90^{+5}_{-11}$ & $M_\rho$  \\
(2000)\cite{CPPACS:mq2:01}& 1-loop TI  &                        &                &                 & ($a\to 0$) \\
QCDSF-UKQCD               & O(a) SW    & $3.5(2)$               & $90^{+5}_{-10}$&                 & $r_0$      \\
(2003)\cite{QCDSF:mq2:00} & 1-loop TI  &                        &                &                 & $[1.9-2.2]$  \\
QCDSF-UKQCD               & O(a) SW    &                        & $85(1)$        &                 & $r_0$      \\
(2003)\cite{QCDSF:mq2:03} & RI-MOM     &                        &                &                 & $[1.9-2.2]$  \\
\hline
\end{tabular}
\caption{Recent $N_f=2$ results for quark masses.}
\label{tab:mq2}
\end{center}
\end{table*}
%

\section{Discussion of $N_f=2$ results}
\label{sec:discuss2}

The CP-PACS collaboration~\cite{CPPACS:mq2:01,CPPACS:mq2:01E} set the stage for large
scale simulations with dynamical fermions by providing results that
are of comparable quality to quenched simulations with respect to
statistics, number of quark masses used in the simulations, and in the
number of values of lattice spacings used in the continuum
extrapolations. Their estimates displayed a number of desired
features. The most striking was that estimates of the strange quark
mass from four different methods $--$ using the axial and vector Ward
identity definition of the quark mass and using either $M_K$ or
$M_\phi$ to fix $m_s$ $--$ were in agreement. The numbers ranged from
$86.9(2.3)$ to $90.3(4.9)$.

{\it The JLQCD collaboration}~\cite{JLQCD:mq2:02} has provided another
measurement of quark masses with two dynamical flavors that
complements results by the CP-PACS collaboration~\cite{CPPACS:mq2:01}.
There are, however, a number of technical differences in the two sets
of calculations. The CP-PACS calculation used the 1-loop mean field
improved value of $c_{SW}$ in the fermion action, whereas the JLQCD
uses the non-perturbative value.  CP-PACS used the Iwasaki improved
gauge action whereas JLQCD uses the unimproved Wilson (plaquette)
action. CP-PACS had results at three values of the lattice scale ($a
\approx 0.22$, $0.16$, and $0.11$ fermi) while JLQCD provide data at a
single point at $a=0.0887(11)$ fermi. (The fourth point in the CP-PACS
calculation at $a=0.0865$ fermi was used only as a consistency check
because it has small statistics.)

\begin{figure}[tbp]  
\begin{center}
\includegraphics[width=0.99\hsize]{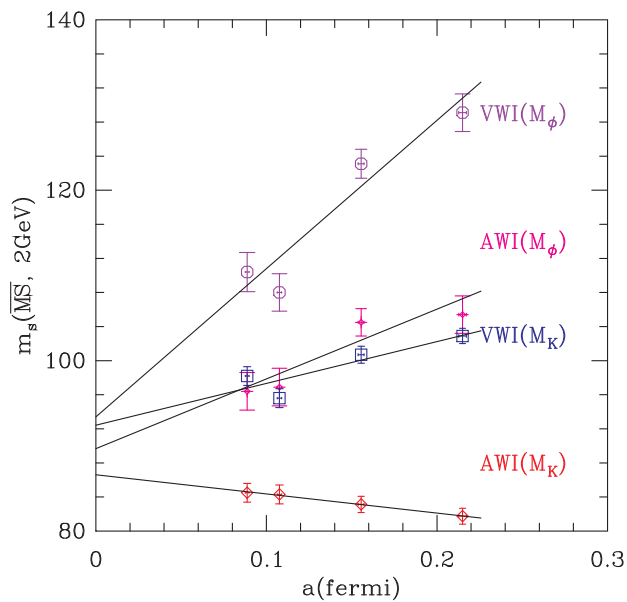}
\caption{Estimates of $m_s$ from the $N_f=2$ simulations by the
CP-PACS and JLQCD (points on the finest lattice with $a \approx 0.09$
fermi) collaborations. Linear extrapolation to the continuum limit are
shown for all four definitions of the quark mass.}
\label{fig:CPPACS_mq}
\end{center}
\end{figure}

The JLQCD estimates spoil some of the nice consistency shown by the
CPPACS analysis. In particular if one combines data from the two
calculations, the extrapolations of AWI($M_K$) and AWI($M_\phi$)
estimates give $\approx 88$ MeV, whereas those from VWI($M_K$) and
VWI($M_\phi$) extrapolate to $\approx 93$ MeV as shown in
Fig.~\ref{fig:CPPACS_mq}. JLQCD quotes their AWI($M_K$) value,
$84.5{}^{+12.0}_{-1.7}$, as their best estimate assuming this method
has the smallest $a$ dependence. The difference between the AWI($M_K$)
and VWI($M_K$) values is taken as an estimate of the systematic
uncertainty. I discuss these data further below.

{\it The QCDSF collaboration}~\cite{QCDSF:mq:03} is in the process of
updating their $N_f=2$ estimate given in~\cite{QCDSF:mq2:00}. The
piece of the calculation still missing is a non-perturbative
evaluation of the renormalization constants. They should be finishing
this calculation soon, meanwhile their unpublished estimate, using
perturbative estimates of renormalization constants, is
$m_s^{\overline {\rm MS}}(2\ {\rm GeV}) = 85(11)$ MeV.

\section{Continuum extrapolation}
\label{sec:context}

I will use the data and fits in Figure~\ref{fig:CPPACS_mq} to
illustrate the systematic uncertainty associated with the continuum
extrapolation and the associated issues of renormalization constants
and the partially quenched approximation. Even though, as mentioned
above, the point at $a \approx 0.09$ fermi is obtained with a
different gauge and fermion action and therefore expected to have a
different coefficient for the $O(a)$ errors, nevertheless, I have
taken the liberty of making a common fit to qualitatively illustrate
how [in]sensitive the conclusions are to current errors in individual
points and to a linear fit that extends all the way to $a = 0.22$
fermi.

The extrapolated values are based on a linear fit to four
points. Keeping a linear term is appropriate since $O(a)$ errors have
not been fully removed from the action or the currents, however this
does not mean that higher order corrections are
unimportant. Looking at the fits it clear that more high precision data
at smaller values of $a$ are required to include/exclude higher order terms 
with a reasonable degree of confidence. Given the spread, JLQCD choose
$84.5{}^{+12.0}_{-1.7}$ as their best estimate of $m_s$ since
$AWI(M_K)$ values show very little $a$ dependence. The different
extrapolations are accommodated by associating a large positive
systematic uncertainty to the central value.

It is interesting to note that the two estimates using $AWI$
extrapolate to $\approx 88$ MeV, whereas those using the $VWI$ to
$\approx 93$ MeV. This suggests that the difference is not due to
using $M_K$ versus $M_\phi$ but due to $AWI$ versus $VWI$ methods.
There are two differences between these methods that could account for
this discrepancy. First, the renormalization constants are different
for lattice actions that do not preserve chiral symmetry (one needs
$Z_A/Z_P$ in the $AWI$ and $Z_m=1/Z_s$ for the $VWI$); and second,
there is an extra complication, in the case of the $VWI$ method,
coming from having to determine $\kappa_c$, the critical value of the
hopping parameter corresponding to zero quark mass. The problem of the
determination of $\kappa_c$ using VWI from partially quenched
simulations leads to an additive shift in estimates of quark
masses. A comparative analysis of $m_{u,d}$ suggests that this
issue leads to no more than $1$ MeV uncertainty in estimate of quark
masses, so I concentrate on the $Z's$ for the explanation.

Repeating the CP-PACS and JLQCD analysis shows that most of the
difference comes from the renormalization factor that connects lattice
results to those in the $\overline {\rm MS}$ scheme at 2 GeV, i.e. $
Z_{VWI} \approx 1.1 Z_{AWI}$.  Another way of stating this is that the
lattice values of $m_s$ are roughly the same for the two methods; it
is the connection between the two schemes which leads to majority of
the difference.  What we know from non-perturbative calculations of
renormalization constants in the quenched approximation is that $Z_P$
is significantly (by about $10\%$) overestimated by tadpole-improved
1-loop perturbation theory, whereas $Z_S$ and $Z_A$ are much better
approximated. If we assume that the same is true in the $N_f=2$ case,
then correcting for this in the CP-PACS/JLQCD data would boost the
$AWI$ results by about $10\%$ since $m_R = (Z_A/Z_P) m$, and 
explain the difference between $AWI$ and $VWI$ results. In that case
all four fits would extrapolate to $m_s(\overline {\rm MS}, {\rm 2\
GeV}) \approx 93$ MeV. In the absence of non-perturbative estimates
for the renormalization constants, my conclusion, based on the CP-PACS
and JLQCD results, is to take a flat distribution between $84-93$ MeV
as the best estimate for $m_s$. This range also incorporates the
QCDSF-UKQCD estimates.

Note that my reservation of $\sim 10\%$ uncertainty due to the 1-loop
$Z$'s does not apply to the four sets of quenched data analyzed by
Wittig because none of those calculations use, simultaneously, the AWI
method and 1-loop estimates for $Z$'s.

\section{$N_f=3$ results}
\label{sec:Nf3}

Simulations with three flavors of dynamical quarks, all with masses
$\le m_s$, represent a qualitatively big step forward. The reason for
this favorable situation is that the coefficients of the chiral
Lagrangian determined by fitting data for observables obtained at
different masses to the corresponding chiral expansions are the same
as QCD~\cite{sharpe:pqqcd:00,sharpe:pqqcd:01,Sharpe:pqqcd:03}.  Thus,
as long as the simulations are done within the region of validity of
$O(p^4)$ $\chi$PT, we can extract physical results even from
simulations done at $3-18$ times $m_d$.

Very recently preliminary results from simulations with three flavors
were reported by the MILC~\cite{Bernard:mq3:03} and
CP-PACS/JLQCD~\cite{JLQCD:mq3:03} collaborations at LATTICE
2003. These are summarized in Table~\ref{tab:mq3}.

\begin{table*}
\begin{center}
\noindent
\setlength{\tabcolsep}{4.2pt}
\begin{tabular}{|l|l|c|c|c|c|}
\hline
                          &  Action    & $\bar m$  & $m_s(M_K)$     & scale $1/a$    \\
                          &  Renorm    &           & AWI            &  (GeV)         \\
\hline						   		     
JLQCD                     & Iwasaki+SW & $2.89(6)$ & $75.6(3.4)$    & $M_\rho$  \\
(2003)\cite{JLQCD:mq3:03} & 1-loop TI  &           &                & ($2.05(5)$)  \\
MILC                      & AsqTad     & $2.5(1)$  & $66(1)$        & $M_\rho$  \\
(2003)\cite{MILC:mq3:03}  & 1-loop TI  &           &                & ($1.6$) \\
MILC                      & AsqTad     & $2.6(1)$  & $68(1)$        & $M_\rho$  \\
(2003)\cite{MILC:mq3:03}  & 1-loop TI  &           &                & ($2.2$) \\
\hline
\end{tabular}
\caption{Recent $N_f=3$ results for quark masses. AsqTad is a
perturbatively improved version of Staggered fermions which reduces
``taste'' symmetry breaking. The quoted errors in the MILC results 
include both statistical and those due to varying
$q^*$ in the 1-loop matching between $(1/a \to 2/a)$. The entries in the 
last column give the quantity used to set the scale and its value in GeV. 
}
\label{tab:mq3}
\end{center}
\end{table*}
%

The CP-PACS/JLQCD collaboration~\cite{JLQCD:mq3:03} employ the same
analysis as in their $N_f=2$ study~\cite{CPPACS:mq2:01} and use an
improved gauge action as well as an $O(a)$ improved Wilson quark
action. Analysis of the AWI data give $m_s = 75.6(3.4)$, and 
the analysis of the VWI data is not complete as the determination of $\kappa_c$
is not yet under control. Their most accurate number is from the AWI($M_K$) method 
and AWI($M_\phi$) gives a consistent value but with much larger errors. The 
difference between the two estimates is folded into the error estimate. 

Two major issues remain with the CP-PACS/JLQCD results. These are (i)
residual discretization errors as the calculation has been done at only one
lattice scale and (ii) the use of 1-loop tadpole improved $Z$'s.  To
address the first requires more data which is a matter of time. On the
second issue my reservation, that the 1-loop perturbation theory
overestimates $Z_P$ by $\sim 10\%$, resurfaces. If this reservation holds
up then their estimate could be as high as $85$ MeV. Based on these
reservations, their numbers suggest the rather large range
$m_s = 70-90$ MeV.

The MILC Collaboration results~\cite{Bernard:mq3:03} are obtained
using improved staggered fermions. (An older estimate, based on an
independent analysis of a sub-set of this MILC data, was reported by
Hein~\cite{Hein:mq3:02} at LATTICE 2002.) I have listed the results
under AWI even though for staggered like fermions (lattice fermions
with a chiral symmetry) the AWI and VWI methods are identical. A major
step forward in the MILC analysis is they perform a combined fit to
data for $M_K^2$ at two sets of lattices (coarse and fine) using a
staggered $\chi$PT expression that includes discretization and taste
symmetry violating corrections
\begin{eqnarray*}
&{}& \hspace{-0.4in} \frac{(M_{K^+_5}^{1-loop})^2}{\mu\,(m_x+m_y)} \nonumber \\
&=& 1 + \frac{1}{16\pi^2f^2} \Big(
[ -\frac{2 a^2 \delta_V^\prime}{M^2_{\eta_V^\prime} - M^2_{\eta_V}}
( l(M^2_{\eta_V}) - l(M^2_{\eta_V^\prime}) ) ]     \nonumber \\
&{}& \hphantom{+++++1} + [ V \to A ] + \frac{2}{3} l(M^2_{\eta_I}) \Big) \nonumber \\
&{}& \hphantom{1} + \frac{16\mu}{f^2}(2L_8-L_5)(m_x+m_y)        \nonumber \\ 
&{}& \hphantom{1} + \frac{32\mu}{f^2}(2L_6-L_4)(2m_q+m_s) \ + \ a^2 C \,,
\end{eqnarray*}
where $m_x+m_y$ is the sum of the masses of the two valence quarks.
Even though fit to a complicated S$\chi$PT expression with 44-46
parameter, having very precise data allows them to extract the central
values and the associated errors estimates reliably. Their best
estimates are $m_{u,d}=2.7(6)$ and $m_s=70(15)$ MeV. A very large part
of the error comes from the fact that to the 1-loop estimate for $Z_m$
they assign an overall $\sim 20\%$ uncertainty due to the neglected
$O(\alpha^2)$ terms.  In my opinion this is a conservative estimate of
the $O(\alpha_s^2)$ uncertainty, especially since the 1-loop
coefficient for the improved (AsqTad) staggered fermions is small, 
($\lsim 0.12\alpha_s$)~\cite{Hein:mq3:02}. The authors are clearly
keeping in mind the lesson learned from quenched unimproved staggered
fermions where that 1-loop perturbation theory underestimated $Z_m$
(and thus $m_s$) by almost $30\%$. 

A concern with the MILC simulation is the lack of a 
``proof'' that the staggered fermion action describes four degenerate
flavors in the continuum limit. Furthermore, there is the potential
problem of loss of locality of the action when taking the square root
and the fourth root of the staggered determinant to simulate two plus
one dynamical flavors. These issues are being
investigated~\cite{Kenchtli:locality:03} now that all other sources of
errors are understood, and the community is moving towards providing
precision results. Unfortunately, as of now there is no airtight
argument that settles these issues.

Based on these two preliminary calculations, and if forced to quote a
single number, my choice is $m_s = 75(15)$ MeV.  This estimate is
certainly very exciting and provocative.  Furthermore, with
simulations at more values of the lattice spacing and with different
fermion formulations coming on line, this exciting result will soon be
refined.

As mentioned before, the power of $N_f=3$ analysis, provided all quark
masses are small such that 1-loop $\chi$PT applies, is that the chiral
parameters are those of physical QCD. Thus, in addition to estimates
for quark masses the MILC collaboration~\cite{Bernard:mq3:03} extract
the Gasser-Leutwyler constants from their fit. In particular they find
that
\begin{equation}
2L_8 - L_5 = - 0.1(1)({}^{+1}_{-3}) \times 10^{-3} \,.
\end{equation}
This is significantly outside the range
\begin{equation}
-3.4\times 10^{-3} \ \lsim\ 2L_8 - L_5 \ \lsim\ -1.8 \times 10^{-3} 
\end{equation}
acceptable for $m_u=0$. The same conclusion has been reached by the OSU
group~\cite{Kilcup:mu3:03}. In short, lattice results do not favor the
possibility that $m_u = 0$ is the solution of the strong CP problem.

\section{$m_s$ from QCD Sum Rules}
\label{sec:sumrules}

Three types of sum-rules have commonly been employed to determine
light quark masses. They are (i) Borel (Laplace) transformed sum rules
(BSR's); (ii) finite energy sum rules (FESR's); and (iii) Hadronic
$\tau$ decay sum rules (these $\tau$-decay SR are a special case of FESR) . 

The starting point for the pseudoscalar and scalar QCD sum rules are
the axial and vector Ward identities 
\begin{eqnarray*}
        \partial^\mu A^{us}_\mu &=&  (m_s+m_u)\ i :\bar s \gamma_5 u: \nonumber \\
        \partial^\mu V^{us}_\mu &=&  (m_s-m_u)\ i :\bar s u:
\end{eqnarray*}
and the corresponding integrated 2-point correlation functions, $e.g.$
\begin{eqnarray*}
&{}& \hspace{-0.4 in} \Psi_{5}(q^2) \\
&\equiv& \hphantom{(m_{d})^2}  i\int\,d^4 x e^{iq\cdot x}
  \langle 0\vert T\{\partial^{\mu}A_{\mu}^{\dag}(x),
  \partial^{\nu}A_{\nu}(0)\}\vert 0\rangle ,  \\
&=& (m_{d}+m_{u})^2            i\int\,d^4 x e^{iq\cdot x}
  \langle 0\vert T\{P^{\dag}(x),P(0)\}\vert 0\rangle \,.
\end{eqnarray*}
$\Psi(q^2)$ and $\Psi_{5}(q^2)$ are analytic on the complex $q^2$
plane with poles and cuts along the positive real axis. They can be
calculated using OPE and perturbation theory for large $|q^2|$ (say $q^2 >
s_0$) and away from the cut.  They are related, through dispersion
relations, to spectral functions, for example, $\rho_5(s) = {\rm Im}
\Psi_5(s)/\pi$.

\begin{figure}
\begin{center}
\includegraphics[height=0.25\vsize]{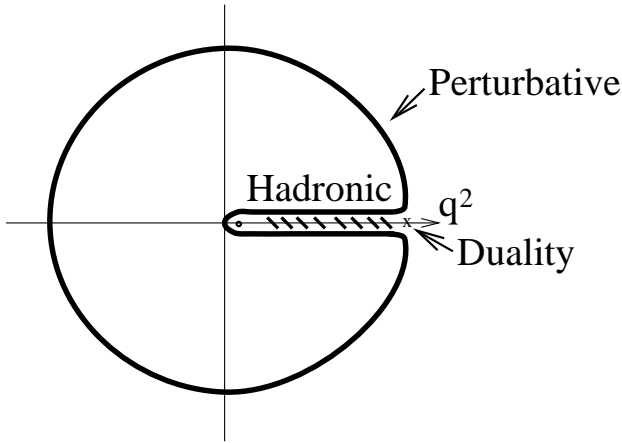}
\caption{The contour integral used in finite energy sum rules. The
poles and cuts are along the real axis. Duality refers to the
matching, on average, between the perturbative and hadronic
ans\"atze.}
\label{fig:contour}
\end{center}
\end{figure}

Finite energy sum rules are based on the observation that the spectral
function has singularities (poles and cuts) only along the real axis
as shown in Fig.~\ref{fig:contour}. The contour integral shown in  Fig.~\ref{fig:contour}
is zero so 
\begin{eqnarray*}
\int_{0}^{s_0} w(s) \rho_{hadronic}\ ds = \frac{-1}{2\pi i} \oint_{|s|=s_0} w(s) \Pi_{PQCD}\ ds
\end{eqnarray*}
The left hand side (discontinuity along the real axis) is evaluated
using a combination of experimental input and modeling for the
spectral function. The right hand side is evaluated using the OPE and
perturbation theory for the dominant mass-dimension $D=0$ term. The scheme and scale
($s_0$) used for the perturbation defines the scheme and scale in
which the quark mass is defined. $w(s)$ are conveniently chosen
weights designed to improve convergence.  Since the OPE is expected to
break down near the real axis at $s_0$, recent analyses have employed
``pinched'' weights like $w(s)=(1 - s/s_0)^n$ that have a zero at
$s=s_0$. The resulting sum rules are called pinched FESR.

In the Borel transformed sum rule one integrates the spectral function 
of the vector current along the real axis
\begin{eqnarray*}
{\cal B} [\Pi]_{OPE}  &=& \hphantom{+} \int_0^{s_0} e^{-s/M^2} \rho_{hadronic} ds \nonumber \\
                      &{}& + \int_{s_0}^\infty e^{-s/M^2} \rho_{OPE} ds \,.
\end{eqnarray*}
For the strange scalar channel the $l.h.s.$ is proportional to $(m_s -
m_u)^2$. In the OPE, the dominant $D=0$ term is evaluated
using perturbation theory and defines the scheme and scale at which
the mass is evaluated. The breakup of the integral on the $r.h.s$
depends on a suitable choice of $s_0$. It has to be large enough that
the integral of the perturbative estimate of the OPE (second term) is
reliable and yet small enough that there is experimental data on the
spectral function up to that point. Otherwise there is a large gap in
which $\rho$ can, at best, be modeled. Second, the answer should be
independent of the Borel Mass $M$. One cannot choose M too small as
the transform gives more weight to the low $s/M^2$ region. This is
good on the $r.h.s.$ but unfortunately on the $l.h.s.$ it enhances the
uncertainty of the higher dimensional operators. At the same time we
want $s_0/M^2 > 1$ so that the unknown contribution of the
``continuum'' to $\rho_{hadronic}$ is suppressed.

There are three important questions central to the reliability of all
sum rules analyses.
\begin{itemize}
\item
How well does the operator product expansion converge? Furthermore,
are the non-perturbative corrections, like quark and gluon
condensates, instanton effects, and neglected higher order terms in
the OPE small?
\item
How well is the perturbative expansion for the leading terms in the
OPE known and how well does it converge at the scale $s_0$?
\item
How well is the hadronic spectral function determined through a combination of 
experimental data and modeling?
\end{itemize}
With respect to these points two major improvements have occurred
over the last five years.  These include
\begin{itemize}
\item
The perturbative series for the scalar and the pseudo-scalar sum-rules
are now known up to $\alpha_s^3$ (four loops)~\cite{Chetyrkin:4loop:97}. 
\item
Better models of the hadronic spectral function have been developed that satisfy 
a number of consistency checks.
\end{itemize}
A number of hurdles, mainly in our ability to determine the phenomenological 
spectral function, remain.
\begin{itemize}
\item
In the pseudoscalar sum rule for $m_u+m_d$, the hadronic spectral
function includes the masses and widths of the kaon, $K(1460)$, and
$K(1830)$ resonances (to extract $m_u+m_d$ from the pion channel the
corresponding states are $\pi, \pi(1300), \pi(1770)$). What are not
known are the decay constants of the $K(1460)$, and $K(1830)$ and
their relative phase.  Also, the $K\pi\pi$ continuum is modeled using
resonant forms with or without chiral perturbation theory
modifications. The prospects of new data to improve the spectral
function are small.
\item
In the scalar sum rule for $m_s$, the spectral function starts at the
$K\pi$ threshold.  Also known are the masses and widths of the
$K_0^\ast(1430) $ and $K_0^\ast(1950) $ resonances. Below the
$K_0^\ast(1430) $ threshold, the spectral function is fairly well
determined using the $K_{e3}$ data and the $K\pi$ scattering
phases. What is not known is whether there is significant phase
variation near and above the $K_0^\ast(1950) $ resonance. Prospects of
improving $\rho(s)$ from B-factories are marginal.
\end{itemize}

Without going into details which will be presented
in~\cite{qm:revBGM:03}, my conclusion, based on the Borel and finite
energy sum rules, is that the most complete analysis for $m_s$ from
the scalar channel gives $m_s= 99(16)$ MeV~\cite{sr:msJOP:01}. For the 
pseudo-scalar channel it is $m_s= 100(12)$~\cite{Maltman:ms:02}.

The $\tau$-decay sum rules utilizes data for the ratio of
semi-hadronic to leptonic decay rate
\begin{eqnarray*}
R_\tau^{V/A,ij} = \frac{\Gamma [ \tau^- \to \nu_\tau {\rm hadrons}_{V/A,ij} (\gamma) ] }
                       {\Gamma [ \tau^- \to \nu_\tau e^- \bar\nu_e(\gamma) ]}
\end{eqnarray*}
where $V/A,ij$ denotes the flavor ($ud$ or $us$) of the Vector $(V)$
or Axial $(A)$ current.  Experimental data gives access to the $u,d$
and $u,s$ spectral functions. The perturbative series for the $1+0$
part of the $D=2$ $\tau$-decay sum rule are known only up to $\alpha_s^2$
(three loops). The $0$ part of the $\tau$-decay series is known to
$\alpha_s^3$. Here $0$ and $1$ refer to the angular momentum of the
hadronic part.  A summary of current estimates of $m_s$ from hadronic
$\tau$-decay sum rules is given in
Table~\ref{tab:taudecay}~\cite{qm:revDPF00:00}.

\begin{table*}[htbp]
\begin{center}
\begin{tabular}{|l| l| l| l|}
\hline
&\multicolumn{3}{c|}{$m_s(2\ {\rm GeV})$ (MeV)}\\
\hline
{Reference}                        &{Original}                         &CKMU input        &CKMN input\\
\hline
CKP98~\cite{tsr:msCKP:98}          &$145\pm 36$ ($|V_{us}|=.2213$)       &$116\pm 31$       &$99\pm 34$\\
KKP00~\cite{tsr:msKKP:00}          &$125\pm 28$ ($|V_{us}|=.2218$)       &$120\pm 28$       &$106\pm 32$\\
KM00~\cite{tsr:msKM:00}            &$115\pm 17$ ($|V_{us}|=.2196$)       &$110\pm 16$       &$100\pm 18$\\
CDGHPP01~\cite{tsr:CDGHPP:01}      &$116^{+20}_{-25}\ \ $ ($|V_{us}|=.2215$) &                  & \\
GPJSP03~\cite{tsr:GPJSP:03}        &                                   &$117\pm 17$       &$103\pm17$\\
\hline
\end{tabular}
\caption{Evolution of results for $m_s(2\ {\rm GeV})$ obtained using
hadronic $\tau$ decay data.  The ``original'' values are those
obtained by the authors along with $|V_{us}|$ used.  The updates of these values use the CKMU
($|V_{us}| = 0.2225\pm0.0021$) and CKMN ($|V_{us}| =
0.2196\pm0.0026$). All CKM[U,N] input sets correspond to $R^{V+A}_{\tau
;us}=.163$ and branching fractions $B_e=.1783$, $B_\mu=.1737$. From
this one gets $R^{V+A}_{\tau ;ud}=3.471$.  All use 
$\alpha_s(m_\tau^2)=.334$ and, with the exception of
KKP00~\cite{tsr:msKKP:00}, the CDGHPP01~\cite{tsr:CDGHPP:01}
truncation prescription.  Experimental and theoretical errors have
been combined in quadrature.}
\vskip .01 in\noindent
\label{tab:taudecay}
\end{center}
\end{table*}

The interesting feature to note, notwithstanding the many
improvements, are that the results in Table~\ref{tab:taudecay} have
been fairly constant over the last three years.  The central value of
$m_s$ has stayed in the range $115-120$ MeV if unitarity of CKM matrix
is imposed, and $m_s = 100-105$ MeV if the Particle Data Group values
for $|V_{us}|$ are used. Both estimates have errors of about $20$ MeV.
In short, the results are very sensitive to the value of $|V_{us}|$,
and the difference between the two estimates is of the same size as all 
the other uncertainties combined. 

Drawbacks of the $\tau$-decay sum rule are: (i) The Cabibbo suppressed
hadronic $\tau$-decay data has not been separately resolved into $J=0$
and $J=1$ contributions and (ii) there is large uncertainty in the
perturbative behavior of the scalar component. Some progress in
reducing a sub-set of these uncertainties was recently reported by the
GPJSP collaboration~\cite{tsr:GPJSP:03} where they used a
phenomenological parameterization for the scalar and pseudoscalar
spectral function in the OPE. These new results, shown in
Table~\ref{tab:taudecay}, are, nevertheless, consistent with previous
estimates.

In terms of future prospects, we expect significant improvement in the
measured $\tau$-decay spectral function, especially above the $K^\ast$.
Meanwhile, it is clear that, at least as far as the central value of 
$m_s$ is concerned, pushing it significantly below $100$ MeV is 
disfavored by the sum-rules analyses.

\section{Lower bounds on quark masses}

Even though the sum rule and Lattice QCD estimates overlap within
combined uncertainties, the current best estimate of the lattice result
($75\pm15$) MeV is tantalizingly small. The following question is
often raised. Does this lattice number violate rigorous lower bounds
predicted from a sum-rule analysis? My answer is NO and I give a brief
justification for this~\cite{qm:revBGM:03}.

The most stringent bound predicted is the ``quadratic'' bound obtained
by Lellouch, de Rafael and Taron~\cite{sr:quadbound:98}. It predicts,
assuming perturbation theory becomes reliable by $Q=2 $ GeV, that
$m_s({\overline {\rm MS}}, 2\ {\rm GeV}) > 100$ MeV.  The Achilles'
heel of this analysis is that the perturbative expression that enters
into the quadratic bound has very large
coefficients~\cite{Maltman:ms:02}:
\begin{eqnarray}
3 {\cal F}_0 {\cal F}_2  - 2 ({\cal F}_1)^2 &=& 1 + \frac{25}{3} a(Q^2) + 61.79 a^2(Q^2) \nonumber \\
                                            &{}& \hphantom{0} +\ 517.15 a^3(Q^2) + \ldots \nonumber \\
                                        &=& 1 + 0.83 + 0.61 + 0.51 + \ldots \,.
\end{eqnarray}
where the second expression has been evaluated at $Q = 2$ GeV with $a
\equiv \alpha_s/\pi \approx 0.1$.  Pushing $Q \ge 2.5$ GeV already
lowers the bound to $m_s({\overline {\rm MS}}, 2\ {\rm GeV}) > 80$ MeV
and going to a still safer value of $Q = 3 $ GeV where $\alpha_s/\pi =
0.086$ gives $m_s({\overline {\rm MS}}, 2\ {\rm GeV}) > 60$ MeV. With
such a poorly behaved series it becomes an article of faith as to what
$Q$ is considered safe, and therefore what value to take as the lower bound.

There are two other bounds for which the perturbation theory is
reasonable even for $Q \ge 1.4$ GeV. These are the $\Sigma_0 \ge 0$
given in~\cite{sr:quadbound:98}, and the ``ratio'' bound given
in~\cite{sr:BGM:98}. Assuming that perturbation theory is reliable at
$Q = 1.4$, both these bounds provide a much lower value, $i.e.$,
$m_s({\overline {\rm MS}}, 2\ {\rm GeV}) > 80$ MeV.

My bottom line on these bounds is that we should be concerned only if
lattice results go significantly below $m_s = 70 MeV$. So, the 
interesting question is which estimate will change over time? Will the
sum rule estimate come down to match the lattice value of $\approx 75$
MeV, or will the lattice result rise to match the sum-rule value $m_s
\gsim 100$ MeV? Or will both change within their errors and come
together in the middle? And finally, it will be interesting to understand
which, if any, systematic error is being underestimated in the two methods.

\bibliography{paper}


\end{document}